\documentclass[12]{article}
\usepackage{mathtext}
\newcommand{\nn}{\medskip} 
\newcommand{\n}{\smallskip} 

\renewcommand{\d}{\delta}
\renewcommand{\b}{\beta}

\newcommand{\QFT}{{\rm QFT}}
\newcommand{\D}{\Delta}

\newcommand{\e}{\epsilon}
\newcommand{\FT}{{\rm FT}}

\newcommand{\ar}{\longrightarrow}
\newcommand{\w}{\omega}
\newcommand{\s}{\sigma}
\newcommand{\la}{\lambda}
\renewcommand{\a}{\alpha}
\begin{document}
\title{Limitation of computational resource as physical principle}
\author{Yuri I. Ozhigov\thanks{e-mail: 
ozhigov@ftian.oivta.ru.} \\[7mm]
Moscow State University,\\
Institute of Physics and Technology,\\
Russian Academy of Sciences,\\ 
} 
\maketitle
\begin{abstract} Limitation of computational resources is considered as a universal principle that for simulation is as fundamental as physical laws are. It claims that all experimentally verifiable implications of physical laws can be simulated by the effective classical algorithms. It is demonstrated through a completely deterministic approach proposed for the simulation of biopolymers assembly. A state of molecule during its assembly is described in terms of the reduced density matrix permitting only limited tunneling. An assembly is treated as a sequence of elementary scatterings of simple molecules from the environment on the point of assembly. A decoherence is treated as a forced measurement of quantum state resulted from the shortage of computational resource. All results of measurements are determined by a choice from the limited number of special options of the nonphysical nature which stay unchanged till the completion of assembly; we do not use the random numbers generators. Observations of equal states during the assembly always give the same result. We treat a scenario of assembly as an establishing of the initial states for elementary scatterings. The different scenarios are compared according to the distributions of the assembly results. 
\end{abstract}
\newpage
\ \ \ \ \ \ \ \ \ \ \ \ \ \ \ \ \ \ \ \ \ \ \ \ \ \ \ \ \ \ \ \ \ \ \ \ \ \ \ \ \ \ \ \ \ \ \ \ \ \ \ \ \ \ \ \ \ \ \ \ \ \ \ \ \ \ \ \ \ \ \ \ \ \ \ \ \ \ \ \ \ \ \ \ \ \ \ \ {\it "God does not play dice"

\ \ \ \ \ \ \ \ \ \ \ \ \ \ \ \ \ \ \ \ \ \ \ \ \ \ \ \ \ \ \ \ \ \ \ \ \ \ \ \ \ \ \ \ \ \ \ \ \ \ \ \ \ \ \ \ \ \ \ \ \ \ \ \ \ \ \ \ \ \ \ \ \ \ \ \ \ \ \ \ \ \ \ \ \ \ \ Albert Einstein}

\section{Introduction}

All experimentally confirmed implications of physical theories can be represented as the outputs of some algorithms that can be called the algorithms simulating physics. No practically important physical predictions can be obtained without using some computational device (before the invention of real computers this role was played by a physicist who performed calculations himself). Hence, such a generalized computer must be considered as a valid physical device. It means that all principal limitations existing for such an abstract device must have the status equivalent to the physical laws. This status is confirmed by the existence of the exact and comprehensive description of such a generalized computer in the framework of the classical algorithm theory which remains unchanged as the computational technique develops (see \cite{Mal, Ro}). Underline that we are dealing with classical algorithms, and moreover - with algorithms which do not require a prohibitively large time and space, that are realizable by real computers at least in principle. Such algorithms are called effective. In the formal algorithm theory we call an algorithm effective if it requires the time limited by some polynomial $p(n)$ to process an input word of the length $n$, and this polynomial does not depend on an input word. Then, say algorithms containing a bruit force are not effective because they require an exponential time. In the quantum physics formalism, too, we can separate the part admitting an effective simulation. All implications of the quantum theory confirmed in experiments belong to this part\footnote{I hope that the reader clearly understands that the simulation of physics by the classical computational methods cannot be treated as the simulation of the quantum physics by the classical physics.}.  

It makes possible to formulate the principle of computational resource limitation as follows: all experimentally verifiable implications of physical laws can be simulated by the effective classical algorithms. But in the end of twentieth century the scheme of experiment was invented that could verify the other part of quantum theory, which in all likelihood does not admit an effective simulation. It is the scheme of quantum computer. Building of a scalable (potentially unlimited) quantum computer does not contradict to the known laws of quantum physics, hence it can be treated as a radical 
experiment verifying quantum theory 
\footnote{Say, the limitation of the number of photons used in quantum dots interactions  does not imply the same limitation of the dimensionality of the computational Hilbert space serving the basis of quantum computations, and it is theoretically possible to have an exponential dimensionality of this space with the comparatively small number of photons.}. 

 If only we can build a quantum computer, then a lot of concrete computational tasks could be solved substantially faster than it is possible by the known computational methods on classical computers. This indicates that no classical device can simulate the work of a quantum computer with a thousand qubits with the same performance by means of known classical methods. Thus the practical realization of a large scale quantum computer would be the unique check of that part of quantum theory which in all probability does not allow an effective (classical) simulation. But it is not the all. We know that in the solution of concrete computational tasks (e.g. that tasks which formulation does not contain an external device - oracle) the advantage of quantum computer is established only comparatively to the known classical algorithms. The question about the existence of classical algorithms as fast as quantum ones remains open despite of the numerous attempts to solve it in negative as well as in positive sense (see, for example, \cite{QC}). In other words in the question of possibility to simulate the physics effectively the mathematics refuses to serve us. It is the bad omen for the future of unlimited quantum computations. It can point out (but does not necessary point!) that the part of physics which does not admit an effective simulation have no attitude to the reality at all. Nevertheless this statement is too fuzzy to be discussed here. Mention only that there are the lot of artificial devices that have no natural analogs and a quantum computer is just of this sort. At the same time it may happen that the natural multi-particle ensembles like molecules of proteins and nuclei acids which play the key role in the functioning of living organisms just allow an effective simulation on classical computers. 

We see that the principle of computational resource limitation cannot be reduced to the physical laws. It can be considered as an independent principle representing an alternative to a scalable quantum computer\footnote{Of course, this alternative is not strict. In all likelihood a limited quantum computer is possible and it could give a valuable advantage over classical ones; it makes sense to build such a computer. Furthermore a classical computer cannot be completely scalable.}. 
We try to derive the method of simulation of complex systems with quantum behavior 
as an assembly of molecules with the linear order from the principle of computational resource limitation. This method may be interesting for the advanced nanotechnology\footnote{I assume that advanced nanotechnology deals with objects of the typical size no more than few, not tens nanometers, for which quantum effects play a crucial role.}. To predict the macroscopic effect of the work of nanostructures we must overcome the barrier of the order $10^8 - 10^{10}$, that separates the typical sizes of these two worlds. For this purpose we must be able to separate the directed evolution from the interfering effect of chaos. Assembly scenarios will be the main objects of our approach. It can be generalized to the evolutions that possess the "time arrow" e.g. the explicit direction. Our approach is the attempt to drug the quantum formalism through the needle hole of limited computational resources which always arise on a way to really complex systems. For this we have to cut the working square for the quantum mechanics introducing forced measurements of entangled quantum states. 

Any formalism applicable to the nanotechnology must satisfy two mutually contradicting conditions: it must be scalable, e.g. it must allow the simple adjunction of new elements, and it must take into account the quantum character of the simulated system. The both conditions require the large resources.  We should learn how to control the deficit computational resource, otherwise we risk to spend all this resource on the calculation of a few quantum dots and do not get to the really interesting functional structures. We need the exact criteria for the computational resource redistribution in course of simulation, and the simulation must be completely performed by a computer. Molecules represent the single example of working nanostructures. Thus any mathematical model of nanostructures at first should be applied to molecules. 

We thus consider a computer as a physical device obeying the classical algorithm theory, in particular possessing the limited computational resource. This assumption implies the serious consequences. For example, we consider a decoherence as a forced measurement of the quantum state in the case when this state is so large that it cannot fit to the classical memory\footnote{By the way, it does not mean that only few particles can participate in quantum superposition in the framework of our model: a state of medium sized fulleren molecule quantumly passing as a whole through the closed narrow slits can be represented by a state $\a|00\ldots 0\rangle +\b|11\ldots ,1\rangle$ and it does not require large memory. It means also that the computational resource cannot be reduced to the degree of entanglement, because this generalized Shroedinger cat is maximally entangled state that requires a small computational resource for its description. Kolmogorov complexity would be the more appropriate characteristic of the simplicity of quantum states description. The reason why I choose computational resource as time+space is that we have to simulate dynamical scenarios rather than objects.}. Moreover, we have to make the following assumption that is absurd from the viewpoint of the "copenhagen interpretation" of quantum mechanics: the different measurements of the same quantum states must give the same result.\footnote{Our approach is alternative also for the many-worlds interpretation of the quantum mechanics of Everett (see. \cite{Ev}). The principle of limitation of the computational resources can be extended to the wide range of quantum processes possessing a scenario, and it gives no contradictions. A quantum computer only could be a single source of such a contradiction (see. Appendix 2 for the details).} 

\section{Model of scenarios}

We outline the essence of the proposed approach. All the following sections are designed for the reader who is interested in the details. 

Let $\Omega$ be a finite alphabet each letter of which corresponds to the small molecular structure which electron terms can be approximately calculated by a method of the Hartree-Fock type. Separate the group of letters $\Omega_0\in\Omega$, denoting assembly elements. Assume that we can approximately solve each scattering problem of every element $a\in\Omega_0$ on any system consisting of no more than three elements of $\Omega$, with the determined spatial configuration. For each structure $a\in\Omega$ let $\rho_a$ denote the set of all possible density matrices corresponding to $a$ 
taking into account its movement in the space and turning. Let $A$ be some finite set and $f:A\ar\Omega$, $h:A\ar R^3$ be some functions. We call a figure of the form
$$
-A_1-A_2-\ldots -A_n-
$$
a chain, where $A_j\in A$, $f(A_j)=a_{i_j}$, $h(A_j)=q_{i_j}, $ $a_{i_j}\in\Omega$ is connected with the $a_{i_{j-1}}$ and with the $a_{i_{j+1}}$ by the covalent electron bounds, $q_{i_j}$ - the coordinates of the structure $a_{i_j}$ in the coordinate system connected with the center of mass of the structure $a_{i_{j-1}}$, when $j>0$, or in the inertial system of the center of  mass when $j=0$.
\footnote{Of course the position of center of mass of $a$ depends on the choice of the density matrix $\rho\in \rho_a$, but we neglect this dependence in order to omit insignificant details. We also assume that $q_{i_j}$ completely determines the form of the electron terms realizing the connection of $a_{i_{j-1}}$ with $a_{i_j}$, in order not to include the description of these electron terms to $q_{i_j}$.} Note that this interpretation of a chain is not the single possible. We could treat $A_j$ as a sequence of electromagnetic impulses of the determined frequencies (see the following section).

An active system is a set of the form 
$$
S:\ \ \ -A_1-A_2-\ldots -A_n-,\ -B_1-B_2-\ldots -B_m-, s, \ l_1,\ldots ,l_k,
$$
where the first two objects are the chains called the coding and the growing respectively, $k>0$, $l_j\ (j=1,2,\ldots ,k)$ - the description of the hydrogen and covalent bounds between $f(A_{s-j+1})$ and $f(B_{m-j+1})$.  Let $R\subseteq\Omega_0$. A state of reservoir with an element of assembly from $R$, that is involved in Brownian motion can be described by the density matrix of the form
\begin{equation}
\rho_{res}=\sum\limits_{a\in R,\ \rho\in \rho_a}p_{\rho}\rho
\label{rho}
\end{equation}
where the probabilities $p$ depend on the concentration of the corresponding elements and on their speed distribution. In the simplest case they can be treated as the same and nonzero only for a few states of assembly elements, for which the probability of the bounding to the growing chain in the scattering is large. Denote by $S_a$ the scattering matrix of the element $a$ on the system AB, formed by the connection between $A_s$ and $B_m$ with the description $l_1$. 
The result of the scattering of the state (\ref{rho}) on AB will be
\begin{equation}
\rho_{scat}=\sum\limits_{a\in R,\ \rho\in \rho_a}p_{\rho}S_{\rho}\rho S_{\rho}^*
\label{disp}
\end{equation}
We call a summand in (\ref{disp}) admitted for assembly if this summand $S_{\rho}\rho S_{\rho}^*$ is a linear combination of the composite systems, each of which consists of AB and some elements $A'$, where $f(A')=a$, $\rho\in \rho_a$, $A'$ is connected with B by the covalent bound, and with A by the covalent or the hydrogen bound. Each of such composite systems corresponds to the new active system whose growing chain is on one unit longer. Given an admitted summand we call the corresponding result of the step of assembly the active system chosen from this linear combination by the special $\theta$- function, such that to extract its value we need also a value of the special option of simulation (see the section 4). 
A sequence of steps of assembly is called a process of assembly. The general number of steps in this process can be limited by some constant $T_0$. The growing chain of active system resulted from the last step of assembly is called the result of this assembly process. It can be convenient to describe the chains classically in the intervals between the nearest instants of scattering, in the Appendix 1 it is shown how to pass to the classical description.

Define a key notion of our approach: a scenario.  
In our definition a scenario is a choice of the density matrix of the reservoir $\rho_{res}$ of the form (\ref{rho}), which, generally speaking can depend on the step of assembly. Roughly speaking it means that we decide beforehand: which element, when and how must be delivered to the point of assembly. Such a program may be called a scenario. If we determine a scenario, a value of the option for $\theta$ - function and an initial state of the active system, then we can obtain the result of assembly by the straightforward computation. Note that $\theta$ -function points to only one basic summand of the linear combination, hence the step of assembly may be determined not for all values of the option. In this case we assume that for the given scenario and option value the assembly is impossible. Imagine now that we are given a sample of chain. We then can compare the result of assembly with this sample and make a conclusion is the chosen option value lucky or not. Searching through all values of the option and computing the result of assembly for each value we can find the fraction of lucky options for this scenario. If this fraction exceeds some limit we can consider this scenario as a successful. We can compare several alternative scenarios of the assembly by this method and choose the best.
\footnote{We shall not take up algorithms for finding the optimal scenario.}
We thus obtain the general prescription how to evaluate a complex process taking into account wide class of quantum effects. Such evaluation cannot be built on the first principles. At last we mention the price that we must pay for it. We must assume that measurements of the same state always give the same result, at least during one passage through the whole assembly process. Just this assumption allows to apply the same option value during the assembly. If it is not so we would fail. Imagine that we assemble the complementary chain of DNA molecule containing of order $10^9$ nucleotides of the four types each of which contains about three tens atoms. We can presuppose the satisfactory precision of the calculation of $S$ matrices for $64$ problems of the scattering, but it is inadmissible to consider the results of the scatterings predetermined. Some small details of the scenario can play a role, for example the presence of some object changing the potential in the vicinity of the point of assembly. If only we refuse from the determination of the result of measurement we meet the search problems with the degree equal to the length of complete chain! Thus the principle of the computational resource limitation implies an unconventional supposition from the viewpoint of the "copenhagen interpretation", which nevertheless must be used in the simulation. 

The growing of chain is not the single process that can be simulated this way. We can consider the different processes, for example comparison of chains including operations like glueing-break, insertion of subchains, which require elementary operations like sliding. The evaluation of such scenarios may be defined similarly to the method defined above. The procedures mentioned are sufficient to the simulation of the main biochemical processes with the complex molecules: a replication of DNA molecules, a transcription from DNA to RNA, a ripening of RNA molecules, a translation from RNA molecules to the peptide protein chain, a spatial conformation of these molecules, their dissociation and a virus insertion. The advantage of this approach is that it allows to account the influence of the wide class of quantum phenomena in the simulation of the above mentioned types of processes that possess the explicit scenario. On the other hand we refuse to simulate a chaos, for example a Brownian motion of identical particles, because they must have the same value of the option and the formal application of our approach to such a system could give at most the repetition of the results of statistical physics by the absolutely inefficient way of the direct calculation of the probability distributions. In other words we abandon the universality of our approach to the real systems. 

\section{Sequences of electromagnetic impulses as coding chains}

We defined the notion of chains first of all for the applications in biochemistry. But the formal scheme proposed above can have the different realization. We can take a sequence of photon beams with the determined frequencies and polarization as a coding chain, where each of these beams will influence to the different types of assembly elements differently. This results in that the elements stronger interacting with photons obtain the more speed and will prevail on the recent level of the growing chain. Suppose for a simplicity that there is one point of assembly: C, and only two types of the elements with the same reactivity with C and with each other, with the same concentration and the same other parameters but one: there exist two energy levels in the spectrum of internal states of the element A: $E_0$ and $E_1$ such that $E_1-E_0=h\w_0$, the corresponding passage with the emitting of photon is allowed, and there are no such levels in the spectrum of B. Let these elements be placed in the reservoir with the same and small concentrations such that these elements practically do not interact with each other. The mean kinetic energy of them is supposed to be the same. What would happen if we light the reservoir by a beam of frequency $\w_0$? In case of photon scattering on an element of each type the probability that this element passes to a state $\Phi$ with the energy $E$ (we start the account of the nonperturbed state $E_0$) will be proportional to $\rho_{type,\ E}$ - the density of energy levels in the vicinity of $E$ for the elements of this type. In case of state with the kinetic energy $E=E_1$ this density for elements of type A 
will be at least twice higher than for B, because in addition to the set corresponding to the continuous spectrum that is common for the both types in the case of type A the similar set will be added which is obtained by the shift $h\w_0$ down, since here the passages with the absorption of photons are possible. The probabilities of other passages will be the same for the both types. Hence the probability that the elements of type A obtain the additional kinetic energy $h\w_0$ will be higher than for type B. Then after the lightening the elements of type A move faster and in view of the weakness of interaction this situation lasts the sufficiently long time for that the probability of the first adjunction of an element of type A to the point of growth is essentially higher. If there exists the same energy gap for type B such that for the other frequency $\w_1$ the roles of types exchange then given a word $S$ of letters A and B we can choose the respective chain of impulses such that the probability of assembly of the chain $S$ will be higher than of each other chain. If the probability of joining an element of the determined type in one step is $l$ times higher that for the other type, the probability of the assembly of the required chain of the length $n$ will be $l^n$ times higher than of each other chain. We thus can influence the scenario of assembly by photons. Alternatively we could consider a sequence of electromagnetic impulses with the definite frequencies and polarization as a coding chain. In this case 
the time would play a role of covalent bounds in the chain. 

\section{Simulation of the result of measurements}

The main consequence of the computational resource limitation principle is the assumption that measurements of the same state in the model must always give the same result. This assumption makes possible to include quantum mechanics to the model, consider it in more details. 

When realizing our plan we meet the following very serious problem of the probability interpretation of a wave function. According to this interpretation the squared module of a wave function is the probability of finding the system in the corresponding state. The possible results of a measurement of system are exactly its basic states corresponding to this measurement and we can call them events. In the deterministic simulation the probability must be obtained as a relative frequency of the corresponding event. For this aim we must consider the so called urn scheme representing a random choice of element from the finite set of elementary events. Each event will be associated with some value of option taken from the finite set of all such values. Since we know the wave function dynamics we could try to divide the evolution into the short steps and at each step repeat the pseudo random procedure generating the probability distribution corresponding to the current wave function value. Then to determine a result of measurement of each state we must know the parameters of this pseudo random procedure which are called the "hidden variables". But the practical implementation of this approach would result in the fast and useless exhaustion of the computational resource. Really, we would have to repeat a random choice at each step in the framework of our urn scheme. Since such choices are independent the number of "hidden variables" would grow with the time which means the complete collapse. Hence it is impossible to associate any sense (including physical) with the "hidden variables". It is the cause of failure of the approaches based on "hidden variables". 

We see that the generation of randomness is the wrong way. We shall follow the other way. Introduce such options of the simulation that will be unchanged during the assembly and that their choice completely determines the result of any measurement of the simulated system. Here the quantum law for the measurement statistics of the form $p_i =|\langle\Psi (t)\ |\ e_i\rangle |^2$ will be obtained from the initial uniform distribution of the option values that is the subject of the search. Let we are given a scenario. In each step of the search we shall fix the current value of option and with this value compute the whole assembly using the scenario. The generation of randomness will thus be completely removed. Because our aim is an effective simulation we must assume that the number of all possible option values is limited from above by some logarithmic function of the number of all particles in the simulated system. The search of all option values is then accessible if the computational resource grows linearly as the size of simulated system grows. 

We proceed with the formal definitions.  
Let $D$ - be some finite set of option values, and $D_0,  D_1 ,\ldots , D_{N-1}$  be nonintersecting subsets of $D$ of the same cardinality. Assume that the measurement of the system with an option value $d\in D_j$ always gives the basic vector $|j\rangle$. Let $2^D$ denote the set of all subsets of $D$. Let $U(N)$ be the set of all unitary operators on $H$, $T_D$ be the set of all permutations on $D$ and $\theta : U (N)\times H\ar T_D$, $\phi :H\ar 2^D$ be some functions. The set 
$$
(\{D_j\ |\ j=0,1,\ldots , N-1\} ,\ \theta , \phi ) 
$$
is called a deterministic model of the dimension $N$, accuracy $\e$ and volume $card(D_1)N$, if the following conditions are satisfied. 

\begin{enumerate}
\item For every vector of state $|\Psi\rangle$ and unitary operator $U$ the following equality is true 
\begin{equation}
\phi (U\ |\Psi\rangle )=\theta (U,|\Psi\rangle )\phi (|\Psi\rangle ). 
\label{corr}
\end{equation}
\item For every vector of state $|\Psi\rangle$ and every $j=0,1,\ldots , N-1$ 
the number $|\langle\Psi | j\rangle |^2$ differs from $card(\phi (|\Psi\rangle )\cap D_j)/card(D_1)$ by no more than $\e$. 
\end{enumerate}

A complete description of the system thus has the form:
$$
(|\Psi\rangle ,\ d),
$$
where $d\in D$ is the option value. 
\n

{\it Informal comment}
\n

Thinking a little we must agree that this definition reflects all that is required from the deterministic description of unitary evolutions: the first property means the correctness of this description, the second means that if the distribution of the option values in the initial state is uniform (it expresses the fact that we do not associate any physical sense with the options) then the distribution of the results of measurement will be quantum.

The following simple theorem takes place.
\newtheorem{Theorem}{Theorem}
\begin{Theorem}For every $\e >0$ and integer $N$ there exists a deterministic model of the dimension $N$, accuracy $\e$ and volume no more than $[1/\e ]+1$. 
\end{Theorem}

Define $D$ as the set of all pairs of integers of the form $(j,k)$, where $j=0,1,2,\ldots , N-1;\ k=1,\ldots , L$. The set $D_j$ consists of all pairs of the form $(j,k),\ k=0,1,\ldots , L$. For the simplicity consider first the case $N=2$, e.g. systems consisting of one qubit only. Its vector of state has the form $|\psi\rangle =\la |0\rangle +\mu |1\rangle$. Define the result of $\phi$ action on a state $|\psi\rangle =\la |0\rangle +\mu |1\rangle$ as the set of such pairs $(j,k)$, that for $j=0\ \ \ $ $k\in\{ 1,2,\ldots , L_0-1\}$, and for $j=1\ \ \ $ $k\in\{ L_0,L_0+1,\ldots ,L\}$, where $L_0 =[L |\la |^2]$. If we take $L=1/\e$, then the point 2 is satisfied. With our definition the values of  function $\phi$ will be the set of pairs of the form 
$$
J_s=\{ (j,k)\ |\ j=0\ \mbox{and}\ k=1,2,\ldots ,s; \ \mbox{or}\ j=1 \ \mbox{è}\ k=s+1,s+2,\ldots , L\} ,
$$ 
such that for every option value $k$ there exists one and only one
 $j$, such that $(j,k)\in J_s$. 

Define the result of $\theta$ action on the pair consisting of a unitary operator $U$ and a state vector $|\Psi\rangle\in H$. The value $\theta(U,|\Psi\rangle )$ must be the permutation on the set of pairs $(j,k)$. Given $j$ and $k$, we define the result of action of this permutation on this pair as the unique pair of the form $(j',k)$, which is contained in the set $\phi (U\ |\Psi\rangle )$. Then the equality \ref{corr}, and the point 1 are satisfied. 

In the case of system of $n$ qubits we must modify the result of action of $\phi$ on the state $\sum\limits_{j=0}^{N-1}\la_j |j\rangle$. Now this result will be defined as the set of such pairs $(j,k)$, that for the constant $j$ the values of $k$ will be exactly the numbers $L_{j-1}+1,L_{j-1}+2,\ldots , L_j$, where $L_j=[L\sum\limits_{p=0}^j|\la_p|^2]$, $j=-1,0,1,\ldots ,N-1$. All the rest will be the same as above. The theorem is proved. 

The value of this theorem is as follows. Any random choice is removed from the model of unitary evolution. The randomness remains only on the stage of the  determining of initial state. It makes possible to compare immediately the initial and output states and to tune the model accordingly with experimental data. 
The fidelity of such a model is determined by to what extent the corresponding urn model can approximate the probability of the results of measurements, e.g., we can take the fidelity as $L/N$. If we want to reduce the number of basic state of the system at hand then it is admissible to reduce the value of model to the same degree. If we fix the fidelity then it is convenient to assume that the result of the measurement replaces the corresponding part of the option value in the memory of our computer.

Consider a particle without spin that in a fixed time instant is characterized by one coordinate $x$ or (alternatively) by impulse $p$. Let these values can vary on the segment $[-L,L]$. Impulse wave function will be connected with the coordinate wave function by the Fourier transform 
$$
\phi(p)=a\int\limits_{-L}^{L}e^{-ipx}\psi(x)dx,
$$
where $a$ is the normalizing factor. Concentration of the particle on impulse $p_0$ in the coordinate representation corresponds to the traveling wave of the form $ae^{ip_0x}$. If we divide the big segment $[-L,L]$ to small parts of the length $\d \xi$ by the points $\xi_i$, then these points can be considered as the value of option of the simulation. If we choose such value $\xi_i$ we can define the coordinate or impulse of this particle as if we perform the corresponding measurement. If we measure impulse we pass to the impulse representation of the wave function $\phi(p)$, and then divide the segment of options $[-L,L]$ into the sequential segments $\Delta^{\phi(p)}_j$. The measurement of impulse gives such value
 $p$, for which $\xi_i\in\Delta_{j(p)}$. When the measurement of impulse or coordinate is not exact we have the uncertainty relation $\d p\ \d x=1$, treated as the dispersion of the possible values of coordinate and impulses, because this relation resulted from the form of wave functions.

Thus the measurable physical value can be extracted from the wave function (taken in the corresponding representation) and the option value only by application of $\theta$- function, that corresponds to the measurement procedure. What would happen with our one dimensional particle if we make an inaccurate measurement of coordinate e.g. we measure only $m<n$ its binary signs? The fidelity of our model preserves if instead of the initial option values we take their approximation with $m_1$ binary signs such that $n_1-m_1=n-m$, where $N=2^{n_1}$, and for this inaccurate model we consider the function $\theta$. The value $x$ of coordinate resulted from its application is the result of inaccurate measurement. The wave function of coordinate resulted from the measurement has the form $\la'_j=\la_{x+j}/d$, where $j=0,1,\ldots,2^{n-m}-1$, $d=\sqrt{\sum\limits_{j=0}^{2^{n-m}-1}|\la_{x+j}|^2}$ - normalizing factor. 
In view of our agreement we loose the first $m_1$ binary signs of option because their place will be occupied by $x$. A model of the same fidelity for this wave function can be obtained if we build the $D$- division for this wave function and the rest $n_1-m_1$ binary signs of the initial option, considering them as the binary signs of new option. We then can measure a value of coordinate again that is equivalent to that firstly we measured it exactly. Alternatively we can measure an impulse of our particle - for this we should pass to the impulse representation of the wave function and consider the corresponding $D$ - division on the new options. Here the product of errors of the measurement of coordinate and impulse will be always  $\d x\ \d p=2^{-n}$, that expresses the uncertainty principle. Here $2^{-n}$ arises in view of that $x$ and $p$ in the binary strings representation are only proportional to the physical coordinate and impulse. 

At first glance we could try to identify the option of simulation with a dual magnitude relatively to a measured magnitude: for example, for a measured coordinate it will be an impulse. But this idea is wrong. We choose a value of option assuming that they will be unchanged as the wave function changes, only the bounds $L_j$ which define the division into subset $D_j$ change. One dimensional disposition of these subsets corresponds to the set of impulses. If we consider two particles then their impulse space are not one dimensional, it will be $S_1\times S_1$, and now we cannot dispose the two particle impulses according to the topology of the disposition of $D_j$, defined in the theorem; on the other hand their disposition corresponding to the topology of $S_1\times S_1$ makes impossible a natural introduction of the options. We see that it is impossible to ascribe any natural physical sense to the options of simulation. 
 
The sense of the options of simulation is that they do not change and remain constant during the simulation of the whole process of assembly. For the processes with the explicit scenario as an assembly or disassembly of molecules we must use the trick of "databases" that is to calculate the result of elementary scattering including the following measurement only once, then store it and use whenever it is required. The main part of the simulating algorithm then acquires the form of the following sequence $Y_1,Y_2,\ldots ,Y_s$, where each procedure $Y_j$ is the sequence of actions:

\begin{enumerate}
\item Calculation of the result of scattering $\psi_j=S|C_{j-1},\tau\rangle$ starting from the classical state $C_{j-1}$ and the current time instant $\tau$, where the scattering matrix $S$ has the form $S=S_1\bigotimes S_2\bigotimes\ldots\bigotimes S_M$, $S_j$ corresponds to possible combinations of the classical states and time instants.

 \item Application of $\theta_{\psi_j}$- function given a wave function $\psi_j$; the result will be $C_j$. 
\end{enumerate}

\section{Establishing and break of covalent and hydrogen bounds as elementary scattering}

An elementary operation on subsystems in the simulating algorithm is an establishing or a break of covalent and hydrogen bounds. The quantum description of this process can be found if we consider it as a process of scattering of complex particles with the possibility of change of their components. The problem of scattering is to calculate the differential section of scattering $d\s$, equal to the fraction of the number of particles scattered into the element of solid angle $d\Omega$ to the total number of all particles entering the zone of interaction. 

Consider the problem of inelastic scattering with the change of components of the particles applied to the separate elementary particles regardless of their identity. Let the potential energy of colliding particles be 
$V$. Let the index $a$ characterize the structures of the reacting particles and their internal states so that without interaction (e.g. at the large distance from the area of interaction) a state of reacting system is described by the eigenstates $\Phi_a$. Let the Hamiltonian of scattering have the form:
$$
H=H_a+V_a=H_b+V_b 
$$
where $H_a$, $H_b$ are the Hamiltonians of the colliding and scattering particles, $V_a$, $V_b$ are the operators of interaction, where $H_a$ and its eigenvalues and eigenvectors have the form:
$$
H_a=H_{a 0}(\xi )-\frac{\nabla^2}{2m_a}, E_a=\frac{h^2k_a^2}{2m_a}+\e_{n_a},
\Phi_a=\phi_{n_a}(\xi )\exp(ik_ar_a),
$$
where $H_{a0}$ is the Hamiltonian corresponding to the internal energy of the particles, $\xi$ is a parameter determining their internal state and in the eigenvalue of energy the summand is separated corresponding to the kinetic energy of the colliding particles, where the analogous relations take place also for the scattering particles (see. \cite{Da}). The reaction is then described by $S$-matrix such that 
$$
\Psi_a(\infty )=S\Phi_a(-\infty ),
$$
and the probability of the passage from a classical configuration
 $a$ to a classical configuration $b$ will be given by equation
$$
P_{ba}=\frac{2\pi}{h}|\langle\Phi_b\ |\ V\ |\Psi_a\rangle |^2 \rho (E_b),
$$
where $\rho (E_b)$ is the density of energy levels of Hamiltonian $H_a$ in the vicinity of $E_b$. Note that for the application of function $\theta$ it is sufficient to know the probabilities of passages $P_{ab}$. Calculation of $S$ - matrix is reduced to the finding of $\Psi_b$. There are no explicit formulas here but there exists the equation for it which is convenient for the sequential approximation in the sense of perturbation theory:
$$
\Psi_a=\Phi_a+(E_a -H_a +i\eta )^{-1}V_a\Psi_a,
$$
where $\eta$ is a small real parameter. This is an integral equation and for its solution the iteration process can be used which results from its form.  

For reactions with many particles for which the exchange processes play a role like electrons, photons and phonons the more detailed way is accepted when the Hamiltonian of interaction $H_a$ is represented in terms of creations and annihilations of identical particles in states which are numbered by index $i,j$. In our case it could be the following representation:
$$
H=H_{0}+H_{nf}+H_{ef}+H_{ep}.
$$
Here $H_0=H_0^0+H_0^1$ is the energy of noninteracting subsystems consisting of the energy of noninteracting particles $H_0^0=\sum\limits_i E_aa_i^+a_i +\sum\limits_i E_bb_i^+b_i +\sum\limits_i E_cc_i^+c_i +\sum\limits_i E_dd_i^+d_i $ (which includes their impulses) and of the energy of internal interactions $H_0^1$ of subsystems which do not result in their decay or absorption - emitting of particles (interaction electrons and nucleus in molecules),

$H_{nf}=\sum\limits_{i,j,k}\la_{i,j,k}a_i^+a_jd_k+\la_{i,j,k}^*a_j^+a_id_k^+$ -
the interaction of nucleus with phonons which is nonzero mainly only for the protons and means separation of proton resulting from the phonon absorption or emission of phonon resulting from the proton adsorption to the system,

$H_{ef}=\sum\limits_{i,j,k}\mu_{i,j,k}b_i^+b_jd_k+\mu_{i,j,k}^*b_j^+b_id_k^+$ - analogous interaction between phonons and electrons,

$H_{ep}=\sum\limits_{i,j,k}\nu_{i,j,k}c_i^+c_jd_k+\nu_{i,j,k}^*c_j^+c_id_k^+$ - analogous interaction between photons and electrons.

Note that we put on here only the main types of interactions explicitly participating in the biochemical metabolism: the first two types participate in the synthesis of ATP and its dissociation, the third - in the photosynthesis; we could also consider the other processes like interaction between photons and nucleus and between photons and phonons. 

Now expanding a scattering matrix into the raw of perturbation theory (see \cite{Li, BS}), for its elements we obtain the following expression
$$
\begin{array}{ll}
\langle f|S|i\rangle &=\langle f|i\rangle+\frac{-i}{h}\int\limits_{-\infty}^{\infty}dt_1\langle f|V(t_1)|i\rangle+
\frac{(-i)2}{h}\int\limits_{-\infty}^{\infty}dt_2\int\limits_{-\infty}^{\infty}\langle f|V(t_1)V(t_2)|i\rangle\theta (t_1-t_2)+\ldots =\\
&\langle f|S_0|i\rangle+\langle f|S_1|i\rangle+\ldots
\end{array}
$$
Such a raw typically is not converging but rather asymptotic e.g., the best approximation results not from the total summing (it leads to the divergence) but from the partial summing up to some fixed summand. This sum is a sum of products of the operators of creation and annihilation corresponding to the time from which the potential $V$ depends. This sum can be represented by the sum of Feynman's diagrams which express elementary conversions of particles (whose amount in this raw can grow unlimitedly). There exists the theorem about renormalization (see \cite{BS}), stating that the deposits of infinite sums of diagrams of the same types $D_1,D_2,\ldots$, generated by one diagram $D$ can be replaced by the deposit of only one diagram  $D$, but for the particles with the appropriately changed charge and mass (the so called quasi-particles). This trick in the computation of sums gives an excellent accordance with the experimentally obtained sections of scattering. It makes possible to limit our considerations by a finite set of input and output states for a scattering in the Fock space of occupation numbers with a finite set of ways of the passages from input to output state but for renormalized particles. For example we can consider two possible results for a given initial state of the scattering problem: the regular  (corresponding to the direction of assembly) and the singular (leading to the other direction). The two diagonal elements of the reduced density matrix corresponding to these results give the simplest wave function to which the $\theta$ - operator can be applied. It gives the next initial state for the new scattering etc. Note that the obtaining of a singular result does not mean the fail of assembly because it can be performed by the different ways. To consider interactions including photons and phonons we must include these particles to the elements of assembly and to deal with the more wide alphabet $\Omega$. Treating the effects connected with the electrical conductivity of chains we can allow the including to the chain the clouds of free electrons which are characterized not by the coordinates of center of mass like ordinary elements but by their density matrix corresponding to the wave function $\Psi$, which has the form of linear combination of Slater determinants (see \cite{De}). We can consider even more exotic constructions like superconducting ensembles of coupled electrons. Our construction remains valid due to its general form. 

\section{What is the difference between simulation and deduction}
  
Unlike conventional tasks of the theoretical physics the simulation is possible not for all systems but only for those which have an explicit evolutionary scenario. But it is not all. Consideration of many particle systems on the long time segments imposes the severe restrictions to the description of elementary scatterings in the simulating algorithm. The possibility arises that effects negligible for one scattering influence seriously to the result of long sequence of repetitions and thus cannot be ignored (for example, nonlinear optic effects). Hence the simulating algorithm must consist of iterations, e.g. it must return to the description of elementary scattering each moment when it is necessary to correct it. The convergence of such a procedure must be ensured by the reliable approximation of the scattering and the complete knowledge about the direction of simulated evolution, for example for the problems of synthesis and dissociation of biomolecules the banks of known reactions should be applied. Note that the problem of exact description of organic chemistry was intensively treated in the meddle of the twentieth century (Hartree-Fock method, methods MO LCAO of Rhutaan, Pople et al., and the method of density functional - see \cite{De, Zu, Sl, DFT}), but we now yet have no the complete description of the dynamical scenarios of the complex molecules behavior let alone the scenario of life. One reason is that these approaches though were not the first principle approaches themselves, nevertheless were based on the deduction of a scenario from the first principles. Usage of the more powerful computers nowadays makes possible to calculate bounds energy and the spatial arrangement of molecules with a few tens of atoms, and very approximately - with a few hundred (see \cite{Ac}), that finds applications in pharmaceutics. The feature of these programs is that their subject is molecules while in our approach it is dynamical scenarios. Our approach thus can be applied to the reconstruction of the biomolecules behavior in time.

Why the analysis of the dynamical scenarios is more useful than the deductive calculations for a separate molecule state? As an example mention the problem of the nuclear RNA molecule spatial conformation after its assembly (transcription). For  sufficiently short segments this conformation can be calculated from the principle of the minimum of energy. But we can hardly meet such a conformation anywhere because in the living cells the conformation of RNA (and the other biopolymers) is conducted by special proteins - chaperons that give it the form convenient for the following transportation and translation. Here is the other reason for the scenario's approach. All advanced method of molecular simulation are semi empirical, e.g. they need an experimental data results from the structural analysis of complex molecules. We have not such a data for the creation of artificial nanostructures (like nanorobots), hence we must relay on the first principles or iterative procedures like the method of self coordinated field or the density functional method. The more complicated structure we consider the more resources are required for the first principle methods and the more role the scenario's approach plays. For the advanced nanotechnology just the scenarios must be the main object for simulation. 

We deal with such deterministic models of evolution with quantum effects that take into account as much quantum mechanics as possible. Its scenario determines the behavior of environment and its options determine the results of measurements.  The aim of this simulation is the search of scenarios with the best fit to  experimental data. The existence of scenario and options cannot be treated as a total determinism. For example, the determined result of measurement can be in the case when all the amplitude is concentrated on some basic state as in quantum computations. Determined values of the option can be regarded as the presence of laws which we cannot formulate at the current time but we agree that such laws exist and reserve the place for them for the best tuning of our model. The general requirements to such class of models. 

\begin{enumerate}
\item  Linear growth of memory when the number of particles in the simulated system grows, and no more than quadratic growth of the time required for simulation in comparison with the real time. 
\item  A model must be determined, e.g. it cannot contain pseudo random numbers generators. 
\item  The scalability treated as the possibility of the simple passage to the more complicated systems and of the including of all known forms of interaction. 
\end{enumerate}

The first requirement says that we want to simulate the dynamics of complex systems on the long time segments. The quadratic growth of the time arises from the Euler method for the solution of differential equation because for the step  $\d t$ the error accumulated on the interval $T$ will be $T\d t$, and if the error must be constant we obtain just the quadratic on $T$ growth of the total time $T/\d t$. The second requirement means that the options of simulation remain unchanged during the time of simulation. It makes possible to compare the input and output states with the different choice of options and thus to compare the different scenarios. It also means that the statistics of distribution of the results is represented in the explicit form and we do not need to generate pseudo random numbers in the framework of such a model\footnote{And even have not such possibility because the generation of randomness directly contradicts to the main principle of the computational resource limitation.}.
 The generation of randomness can be used if wanted only for the aim of demonstration. The third requirement means that we can easily include all known effects to the model in order to the best tuning of model. Eventually, all these requirements make sense only when a model correctly predicts the result, hence it is useless to evaluate them theoretically. 

The model proposed in this work certainly satisfies the condition 2, and with some limitations - the condition 3. Formally it is possible to ensure the condition 1 but the adequacy of such a model will depend on the limitation on the computational resource which is unknown. If we make the first condition weaker permitting a polynomial growth of the resource then we obtain the class of effective models. Here the effectiveness is a property from the algorithm theory but not a practical characteristic. In Appendix 2 we represent some reasons for that such effectiveness requires the limitation on the complexity of the entangled quantum states. In Appendix 1 it is shown why the passage from the classical description to the quantum and vice versa can be profitable from the viewpoint of the economy of computational resource. In Appendix 3 we show an example of the economy of computational resource with the help of "database" method for the solution of Shroedinger equation. 

\section{Conclusion}

We assume the principle of computational resource limitation. It means that the simulation of 
quantum systems must be performed by effective classical algorithms and must be based on
strict causality. It was illustrated by a model designed first of all for the description of the assembly, dissociation and spatial conformation of linear biomolecules. The main steps of such a model: 

1. There are permitted only superpositions of the limited number of basic states. Quantum unitary evolution is simulated by the reduced density matrix embracing the limited numbers of parameters of the whole molecule. A unitary evolution is represented as an elementary scatterings with the possible change of the chemical structure of participants. 

2. The complete evolution is simulated as a sequence of elementary scatterings. The participants of scatterings from the environment are defined completely by the choice of evolutionary scenario. 

3. The superpositions arisen after the scattering which size is too large to fit into the reduced density matrix are the subject of forced measurement which result is defined beforehand by the chosen options of simulation. These options remain unchanged up to the end of assembly. Measurements of the same states always give the same result during one pass of assembly. 

4. Each resulting system is placed to the database of the results, we choose the following value of options and the pass of assembly is repeated. After the search of all option values we calculate the probability of the different resulting systems for the given scenario.

5. We choose a scenario which gives the better agreement with the observed macroscopic effects. 

There is no randomness in this model. The simulation of the different measurement of the same state always gives the same result if the value of options is the same. If we fix the options then every difference in measurements must have the classically expressed cause, for example the presence of a certain participant in the scattering, the presence of some molecule in its vicinity etc. The effectiveness of such a simulation is determined by the right choice of scenario that for the problem of assembly requires application of molecular structures and reactions databases and also the correct calculation of amplitudes of the scatterings outcomes.  

An evolutionary scenario determined by the classical terms must be the key object of simulation. An essence of simulation is a comparison of scenarios by the probability distributions of their results. The main advantage of our approach is the possibility to take into account all the known quantum effects through the simple including them to the model.

{\large \bf Appendix 1. \ \ Passages between quantum and classical descriptions as a result of the computational resources deficiency}
\nn

Limitation of computational resource can give a criterion of when the simulated system must be described in terms of quantum mechanics and when classically. A principal drawback of classical description is a instability of its trajectories, e.g. their sensitivity to a small variation of initial conditions that takes place for the valuable part of many body problems. Quantum mechanics is free from this disadvantage: in view of the unitary property of evolutions a small variation of initial conditions gives the small divergence of the trajectory in any time instant in future. Consider the typical problem about a passage of a particle through two close slits in a nontransparent screen disposed at the distance $d$ one from the other, and denote their coordinates by $|0\rangle$ and $|1\rangle$. As usual we assume that the passed particles are measured by the detecting screen disposed behind the slits. A classical consideration requires  exact initial conditions of impacting particles because it determines a slit they pass through. Even if $d$ is a constant the computational resource required for the correct solution of this task will converge to the infinity because if a particle impacts to the vicinity of the middle point between slits the required resolution will grow unlimitedly. In the quantum case we can manage with the limited resolution and obtain the approximate statistical picture of the passed particles impacting to the detecting screen. It is possible mainly because a state of passing particle is represented not by its coordinate as in classical case but by the linear combination of the two possible ways of  particle: $\la |0\rangle +\mu |1\rangle$. 

Consider now the reverse situation when the economy of computational resources leads to the passage from the quantum description to the classical. It takes place where the classical trajectory is stable, in such cases a classical description does not require the growth of resolution and thus gives the saving in memory. Consider one dimensional movement of a particle with unit mass in a field with potential $V(t,X)$, where $t$ is the time, $X$ is the coordinate. Denote its impulse by $P$. The Hamiltonian of such a particle has the form
$H=\frac{P^2}{2}+V(t,X)$ in classical as well as in quantum case only in the last case $P=\frac{1}{i}\frac{\partial}{\partial X}$. We give the quantum description of this movement in the form of simulation on a quantum computer that was proposed in \cite{Za, Wi}, and which is in our case equivalent to the straightforward solution of the Cauchy problem for Shroedinger equation by the method of finite differences. This problem has the form:
\begin{equation}
i\frac{\partial\Psi}{\partial t}=\frac{P^2}{2}+V(t,X),\ \ \ \ \Psi(0)=\Psi_0.
\label{wave}
\end{equation}
Its solution is given by the formula $\Psi(t,X)=\exp(-i\int\limits_0^tH(t,X)dt)\Psi_0$. The method of quantum computing simulation is as follows. 

We approximate the action of operator $e^{-iHt}$ on a wave function $\psi_0$ where $H=H_p +H_q$, $H_p =\frac{p^2}{2m}$, $H_q =V(q)$, $p=\frac{1}{i}\frac{\partial}{\partial q}$ and a potential $V(q)$ is a real function. Without loss in generality we can take $t=1$. To obtain a useful approximation we must work in coordinate or in impulse basis in the space of state vectors where in the both cases our Hamiltonian is not diagonal. To reduce the situation to the a simple diagonal form we choose a small segment of time $\D t$ and represent our operator of evolution approximately as
\begin{equation} e^{-iH}\approx (e^{-iH_q \D t}\ e^{-iH_p \D t})^{1/\D t}. 
\label{Ev} 
\end{equation} 
Choosing, say the coordinate basis we obtain the diagonal operator $H_q$. Applying the Fourier transform $\FT :\ f\ar\frac{1}{\sqrt{2\pi}}\int_{-\infty}^{+\infty}e^{-ipq}f(q)\ dq$ and its property to replace the derivative $\partial /\partial\ q$ by the factor $ip$ we can represent the action of impulse part of Hamiltonian as $e^{-iH_p}=\FT^{-1}\ e^{-ip^2 \D t/2m}\ \FT$, where the middle operator is diagonal. If we can perform Fourier transform and the phase shift on $-p^2 /2m$, then the sequential application of these operators from (\ref{Ev}) gives the required approximation. Let a wave function $\psi (q)$ be defined on a segment $(-A,A)$ and its impulse representation $\FT\ \psi$ be defined on a segment $(-B,B)$. Choosing small values $\D q$ and $\D p$, we can approximate it by $\sum\limits_{a=0}^{2A/\D q} \psi (q_a)\d_a$ where $\d_a (q)$ takes the value 1 on the segment $(q_a ,q_a +\D q )$ and 0 for the other $q$. We can then approximate Fourier transform by a linear operator which action on $\d_a$ gives $\frac{1}{\sqrt{2\pi}}\D q\sum\limits_{b=0}^{2B/\D p} e^{-ip_b q_a}\s_b (p)$ where $\s_b (p)$ is an impulse function which is defined by analogy to $\d_a$. Introducing the new functions of such a sort for coordinates and impulses by $d_a (q)=\d_a (q-A)$, $s_b (p)=\s_b (p-B)$, we rewrite Fourier transform as follows
 \begin{equation} d_a \ \ar\ \frac{1}{\sqrt{2\pi}}\D q\sum\limits_{b=0}^{2B/\D p} e^{-i\ ba\D q\ \D p} s_b 
\label{FT} 
\end{equation} 
that looks like quantum Fourier transform. Assume that the physical space is grained in coordinate as well as in impulse representation with the size of grains $\D q$ and $\D p$ respectively. The particle can be in points of the form $q_a$ only, or can have impulses of the form $p_b$ only. We associate a position $q_a$; $\ a=0,1,\ldots , N=2^l$ with the basic state $|a\rangle$ of $\ \ l$ qubit quantum system. For a simplicity choose such a unit of length that 
 $\D q=\D p =\sqrt{2\pi}/\sqrt{N}$ and let $A=B=\sqrt{\pi N/2}$. Then (\ref{FT}) corresponds to quantum Fourier transform of the form 
$$
 \QFT :\ |a\rangle\ar\frac{1}{\sqrt{N}}\sum\limits_{b=0}^{N-1}e^{-\frac{2\pi i\ ab}{N}}|b\rangle , 
$$ 
and the phase shift to $-p^2 \D t/2m $ from (\ref{Ev}) corresponds to the phase shift to $-\pi b^2 \D t /mN $. In the framework of our quantization of the problem the reverse Fourier transform corresponds to the reverse of quantum Fourier transform. Simulating of $s_1$ particle system we should store $s_1$ copies of the quantum register for one particle and perform the above mentioned procedure to obtain the joint quantum memory. 

Similarly we can consider the case of turning of the solid state around some point (typically - around its center of mass). Here the Hamiltonian has the form
$$
H^1=H^1_{kin}+V(\mu ,\nu ,\eta ),\ \ \ H^1_{kin}(\mu ,\nu ,\eta )=\frac{J^2_x}{2I_x}+\frac{J^2_y}{2I_y}+\frac{J^2_z}{2I_z}
$$
where the potential and kinetic energy of turning are expressed through Euler angles $\mu ,\nu ,\eta$ which determine a position of the solid turning around one point, $I$ denotes the moment of inertia corresponding to a given axe, the operator of moment $\bar J=[\bar r\times \bar p]$, where $r$ and $p$ - are the operators of coordinate and impulse. 

Considering an assembly of linear molecule we represented a step of simulating algorithm is an elementary scattering of inelastic type in which not only internal energy can change but also the chemical contents of the reacting particles. But in the intervals between scattering we can consider the motion of molecule classically. How to obtain this description from the quantum? The conventional way of doing so is to apply Erenfest theorem. We now show it straightforwardly. 

The operator of evolution for every $j$-th segment of the molecule has the form $U=\exp{-iH\d t}$, where the Hamiltonian  $H_j=H_j^0+H_j^1$ consists of the summand $H_j^0=\frac{P_j^2}{2m_j}+V(s_j,X_j)$, corresponding to the motion $X_j^0$, and $H_j^1$ corresponding to the turning of the $j$-th segment of the molecule as a whole around the point $X_j$ (it can be chosen as the center of mass of $j$-th segment). The last motion is considered analogously to the motion of a pointwise particle, so we restrict our consideration by this last case. 

By the quantization all impulses and coordinates are represented by the binary integers $x$ and $p$ from interval $[0,1)$, that are connected with the physical magnitudes by the following equations: $P=\sqrt{2\pi /N}p$, $X=\sqrt{2\pi /N}x$. We assume that the derivative of potential $V'(X)$ changes slowly, namely in the neighboring coordinate points this derivative can be considered as constant.  
The sequence corresponding to $H_p$, has the form 
$$
x\ar\sum\limits_pe^{\frac{-2\pi i\ xp}{N}}|p\rangle\ar\frac{1}{\sqrt{N}}\sum\limits_pe^{\frac{-2\pi i\ xp}{N}}e^{iP^2\d t\pi /N}|p\rangle
$$

Let $\theta_p(\kappa )=a$. It means that we must replace $P$ by $P+A$ in the amplitudes of preceding expression, $A= \sqrt{2\pi /N}a$. It gives after the following Fourier transform:
$$
\frac{1}{N}\sum\limits_y\la_y|y\rangle
,\ \ \ \ \la_y=\sum\limits_pe^{ip^2\d t\pi /N}e^{-2\pi i\ xa/N}e^{\frac{ip(-2\pi X+2\pi A\d t+2\pi Y)}{N}}|y\rangle
$$

The first two factors in the expression of amplitude have no physical sense,  because they give the common phase shift. We have $\la_y=0$ for all $y$ but one which corresponds to the current location of the particle $Y=X-A\d t$, e.g. the instantaneous speed will be $P$, which means that one of two main equations of classical mechanics is satisfied: $X'=\partial H/\partial P$. To obtain the other equation consider the passage corresponding to the application of the coordinate part of Hamiltonian: $H_x=V(X)$. We start with the state 
$|p\rangle$. An application of operator $\QFT\ H_x\ \QFT^{-1}$ then gives such a sequence
$$
|p\rangle\ar\frac{1}{\sqrt{N}}\sum\limits_xe^{\frac{2\pi ixp}{N}}|x\rangle\ar\frac{1}{\sqrt{N}}\sum\limits_xe^{\frac{2\pi ixp}{N}}e^{iv(x)}|x\rangle\ar\frac{1}{N}\sum\limits_{p_1}\la_{p_1}|p_1\rangle ,
$$
that 
$$
\la_{p_1}=\sum\limits_xe^{2\pi ix(\frac{p}{N}+\frac{v(x)\d t}{2\pi x}-\frac{p_1}{N}}.
$$
Then $\la_{p_1}\neq 0$ is if and only if the following equation is satisfied 
$$
P_1=P+\d t\frac{V(X)}{X},
$$
that in view of our agreement about the slow change of the potential gradient gives the second equation of classical dynamics:
$$
P'=\frac{\partial H}{\partial x}.
$$

{\large \bf Appendix 2. Limitation of computational resource in quantum computations as the necessary condition of effectiveness of the model}
\nn

In the algorithm theory an effectiveness means a polynomial complexity, e.g. the time of the work of simulating algorithm must be bounded above by a polynomial of the length of input word. A weaker limitation is an effective description of the system e.g., polynomial upper bound of number of the option values. An effective description is not excluded even in the case when there is no effective prediction of the result of measurements. Consider the Hilbert $N=2^n$ dimensional space of states of $n$ qubit system spanned by the basic vectors of the form $|\e_1 , \e_2 , \ldots , \e_n\rangle$, $\e_j\in\{ 0,1\}$. To describe the result of measurement it would suffice to have $n$ bits only. 
We must keep somewhere $N$ amplitude to describe a state and it requires an exponential size of memory. Nevertheless it does not imply that a classical device with a polynomial resource of the time and memory cannot predict the results of measurements so that their statistics will coincide with the quantum mechanical. 

Try to define an effective model. We shall tell apart an evolution without observers and a process of measurement. Suppose that the complete description of  state requires the knowledge of some options $\Omega$ in addition to the conventional wave function. These options together with the wave function $\Psi$ form the complete description of state which we denote by $\tilde\Psi$, and which determines the result of any measurement of this state. If there exist some unknown options of the external fields which influence to the change of this description beyond the potentials, then we cannot use such a "model" practically in principle. Hence, to make our consideration useful we must suppose that the time evolution of this extended state vector $\tilde\Psi$ without an observer is determined by the Hamiltonian of the system only. The Hamiltonian can be described in finite terms using the notion of a quantum gate array and this method is universal because any quantum evolution can be represented in this manner. Namely, associate with each quantum $n$ particle system the quantum gate array implementing the action of its Hamiltonian in the short time interval $\d t$. This array can be then encoded as the Boolean string 
$Ham$, which length grows linearly of $n$, if we assume that an interaction exists between the neighboring particles only. The length of economical code of our system varies like $\log (\d t /\d t_2)$ as the time varies, because we have to spend a logarithmic amount of memory to remember the additional elementary operations. The resulting complete state of the system before the Hamiltonian action with the changed options is expressed by an extended vector of the final state $\tilde\Psi_t$, which is uniquelly determined by the knowledge of the initial state $\tilde\Psi$ and the code of evolution $Ham$. 

The definition of effective model then acquires the following form. 

\begin{enumerate}
\item A) The result of measurement of every state $\tilde\Psi_t$ can be obtained from $\tilde\Psi$ and $Ham$ by some algorithm $A$ of polynomial complexity. 
\item B) When a quantum state with the same wave function  $\tilde\Psi$ is prepared repeatedly its option must be distributed uniformly on the set of all their value. Here the statistics of the measurements results obtained by the algorithm $A$, coincides with the quantum statistics. 
\end{enumerate}

The first condition says that the complexity of computation of the measurement result is not too large. The second means that the options of simulation have no physical sense and we cannot know how to choose them to predict the exact result of measurement. We could permit also not uniform distributions but with the condition that the option values are generated by an algorithm with a polynomial complexity, it would be equivalent. These two conditions look natural and nothing forbids immediately the existence of effective model. An exponentially growing dimensionality of the Hilbert space of states cannot forbid directly the effective models in quantum physics because there are numerous approximate methods reducing this dimensionality: the method of quasi-particles, molecular orbitals etc. But we now represent a very serious argument for such prohibition. It requires the application of relatively complex mathematical results, namely the quantum computing. 

If we assume that there exist problems without oracles which can be solved by a quantum computer more than as polynomial faster than by any classical, then the existence of effective models is incompatible with quantum mechanics. It is worth noting that almost exponential quantum speedup is reached only in comparison to the known classical algorithms, like in the problem of integer factoring. It is yet not proved that it is impossible at all to solve such problems classically with the same efficiency. 

It is interesting that the open mathematical problems is a stumbling block for the question of the possibility to build an effective simulation of physics. Nevertheless we can rely to the experience of numerous and unsuccessful attempts to build classical algorithms essentially more effective than the known ones, and suppose that such fast classical algorithms do not exist. Then for such problems a quantum computer really exceeds every classical computer in speed more than in a polynomial times. For example, assume that for the problem of factorization integers Shor factoring quantum algorithm possesses this effectiveness. 
Using additional qubits - ancillas we can assume that a final result (nontrivial factors of an initial integer) is obtained from the result of measurement with the probability closed to 1. Suppose that there exists an effective model of quantum mechanics. Consider a quantum gate array implementing Shor algorithm. Choose randomly an option values from the uniform distribution, obtaining the initial state of the computer. Apply then to this state and the code of gate array the algorithm of prediction from the point A) which gives the result of final measurement in Shor algorithm. By the point B) the statistics of measurements for the predicting algorithm coincide to quantum statistics, hence the probability to obtain the right result is closed to 1 and we obtain a fast classical probabilistic algorithm for the problem of factoring that contradicts to our initial supposition.
An effective simulation in the world of unlimited Hilbert spaces is thus hardly possible
\footnote{If only not to assume the probability of the breakthrough in mathematics ! But the consideration of such probabilities is not completely constructive: something must be taken on trust.}.
In other words, quantum algorithms represented the formal manipulations with the Hilbert space of states allowed by quantum physics do not permit an effective prediction of the results of measurements, e.g. for satisfaction of the condition 1 for models it is necessary to impose the limitations on the complexity of entangled states that is done in our approach by the reduction of density matrix. 
\nn

{\large \bf Appendix 3. Database for solution of Shroedinger equation}

Consider how a trick "database" can be applied for the solution of Shroedinger equation. It is applicable for a many particle problem of scattering as well. We illustrate it on the equation of the form (\ref{wave}). Choose some quantization of this equation, say a quantum method of Zalka - Wiesner, described above (we could take the conventional method of finite differences, the effect would be the same).  
Applying sequentially the formula (\ref{Ev}), we obtain the solution of Cauchy problem for our equation in the symbolic form:
$$
\sum\limits_{j=0}^{N-1}A(t)_j V_j+ \sum\limits_{j=0}^{N-1}B_j P_j,
$$
where $V_j$ and $P_j$ - are a value of potential and an initial value of wave function in the nodes. For the massive $A(t)$ we obtain a recurrent equation which makes possible to fill this database sequentially for all values  $t=1,2,\ldots , T$. Let we are given a value $T$ (it is typical for such problems as scattering). We then can find the massive $A(T)$ only once, and in future use it for the solution of Shroedinger equation with any potential and initial conditions simply extracting the current values $A(T)_j$ from the memory. 

\end{document}